\documentclass[aip,showpacs,superscriptaddress,onecolumn,preprint]{revtex4}%
\usepackage{amsfonts}
\usepackage{amsmath}
\usepackage{amssymb}
\usepackage{graphicx}%
\begin{document}
\title{\emph{Ab Initio} Simulations of Dense Helium Plasmas}
\author{Cong Wang}
\affiliation{LCP, Institute of Applied Physics and Computational Mathematics, P.O. Box
8009, Beijing 100088, People's Republic of China}
\author{Xian-Tu He}
\affiliation{LCP, Institute of Applied Physics and Computational Mathematics, P.O. Box
8009, Beijing 100088, People's Republic of China}
\affiliation{Center for Applied Physics and Technology, Peking University, Beijing 100871,
People's Republic of China}
\author{Ping Zhang}
\thanks{Corresponding author: zhang\underline{ }ping@iapcm.ac.cn}
\affiliation{LCP, Institute of Applied Physics and Computational Mathematics, P.O. Box
8009, Beijing 100088, People's Republic of China}
\affiliation{Center for Applied Physics and Technology, Peking University, Beijing 100871,
People's Republic of China}

\begin{abstract}
We study the thermophysical properties of dense helium plasmas by
using quantum molecular dynamics and orbital-free molecular dynamics
simulations, where densities are considered from 400 to 800
g/cm$^{3}$ and temperatures up to 800 eV. Results are presented for
the equation of state. From the Kubo-Greenwood formula, we derive
the electrical conductivity and electronic thermal conductivity. In
particular, with the increase in temperature, we discuss the change
in the Lorenz number, which indicates a transition from strong
coupling and degenerate state to moderate coupling and partial
degeneracy regime for dense helium.

\end{abstract}

\pacs{51.30.+i, 52.25.Fi, 52.27.Gr, 52.65.Yy}
\maketitle

Pressure induced physical properties of hot dense helium plasmas are
of crucial interest for inertial confinement fusion (ICF) and
astrophysics
\cite{PBX:Atzeni:2004,PBX:Lindl:1998,PBX:Nuckolls:1972}. Complete
burning of deuterium-tritium (DT) capsule follows the fusion
reaction D+T$\rightarrow$ $n$+He+17.6MeV, thus accurate knowledge of
the equation of states (EOS) and the relative transport coefficients
for helium are essential in typical ICF designs. In the direct-drive
scheme, thermal transports in ICF plasmas play a central role in
predicting laser absorption \cite{PBX:Seka:2008}, shock timing
\cite{PBX:Boehly:2006}, and Rayleigh-Taylor instabilities grown at
the fuel-ablator interface or at the hot spot-fuel interface
\cite{PBX:Azechi:1997,PBX:Glendinning:1997,PBX:Lobatchev:2000},
while in the indirect-drive ICF, the efficiency of the x-ray
conversion is also determined by thermal conduction
\cite{PBX:Lindl:2004}. Accurate modeling of electrical conductivity
is important for precisely determining interactions between
electrons and plasmas, because high-energy electron beam has been
considered to be the most suitable source for igniting the hot spot
much smaller than the dense DT core in the fast ignitor
\cite{PBX:Cai:2009}. Because of these important items mentioned
above, the EOS and electronic transport properties for hot dense
helium are highly recommended to be presented and understood. From
the theoretical point of view, various assumptions have been used to
predict the electrical conductivity and electronic thermal
conductivity for weakly coupled and strongly degenerated plasmas
\cite{PBX:Spitzer:1953,PBX:Hubbard:1966,PBX:Brysk:1975,PBX:Lee:1984}.
These classic methods widely disagree at high density and are
incorrect to model nonlinear screening caused by electronic
polarization. Quantum molecular dynamic (QMD) simulation
\cite{PBX:Kietzmann:2008,PBX:Lorenzen:2009}, which is free of
adjustable parameters or empirical interionic potentials, has been
proven to be ideally suited for studying warm dense matter. However,
large number of occupied electronic states and overlap between
pseudocores limit the use of this method at high temperatures and
high densities.

In the present work, direct QMD simulations based on plane-wave
density functional theory (DFT) have been adopted to study helium at
both high temperatures and high densities, along the 400 to 800
g/cm$^{3}$ isochore and temperatures up to 800 eV. The EOS data have
been determined for a wide range of densities and temperatures. We
apply Kubo-Greenwood formula as a starting point for the evaluation
of the dynamic conductivity $\sigma(\omega)$ from which the dc
conductivity ($\sigma_{dc}$) and electronic thermal conductivity
($K$) can be extracted.

We introduce the \emph{ab initio} plane-wave code ABINIT
\cite{PBX:abinit,PBX:Gonze:2002,PBX:Bottin:2008} to perform QMD
simulations. A series of volume-fixed supercells including $N$
atoms, which are repeated periodically throughout the space, form
the elements of our calculations. After Born-Oppenheimer
approximation, electrons are quantum mechanically treated through
plane-wave, finite-temperature DFT, where the electronic states are
populated according to Fermi-Dirac distributions. The
exchange-correlation functional is determined by local density
approximation (LDA) with Teter-Pade parametrization
\cite{PBX:Goedecker:1996}, and the temperature dependence of
exchange-correlation functional, which is convinced to be as small
as negligible, is not taken into account. The selection of
pseudopotential approximations, which separate core electrons from
valence electrons, prevents general QMD simulations from
high-density region with the convenience of saving computational
cost. Because of the density-induced overlap of the pseudopotential
cutoff radius and delocalization of core electrons, this frozen-core
approximation only works at moderate densities. As a consequence, a
Coulombic pseudopotential with a cutoff radius $r_{s}$ = 0.001 a.u.,
which is used to model dense helium up to 800 g/cm$^{3}$, has been
built to overcome the limitations. The plane-wave cutoff energy is
set to 200.0 a.u., because large basis set is necessary in modelling
wavefunctions near the core. Sufficient occupational band numbers
are included in the overall calculations (the occupation down to
10$^{-6}$ for electronic states are considered). $\Gamma$ point and
3$\times$3$\times$3 Monkhorst-Pack scheme \textbf{k} points are used
to sample the Brillouin zone in molecular dynamics simulations and
electronic structure calculations, respectively, because EOS
(transport coefficients) can only be modified within 5\% (15\%) for
the selection of higher number of \textbf{k} points. A total number
of 64 helium atoms are used in the cubic box. Isokinetic ensemble is
adopted in present simulations, and local equilibrium is kept
through setting the electronic ($T_{e}$) and ionic ($T_{i}$)
temperatures to be equal. Each dynamics simulation is lasted for
6000 steps, and the time steps for the integrations of atomic motion
are selected according to different densities (temperatures)
\cite{PBX:timestep}. Then, EOS is averaged over the subsequent 1000
step simulations.

QMD simulations have only been run for temperatures up to 300 eV,
for higher temperatures, the thermal excited electronic states
increase dramatically, and are currently numerically intractable.
For temperatures between 10 eV and 1000 eV, the EOS of dense helium
are also obtained from orbital-free molecular dynamic (OFMD)
simulations for comparison with the QMD results. In this scheme,
orbital-free functional is derived from the semiclassical
development of the Mermin functional \cite{PBX:Brack:2003}, which
leads to the finite-temperature Thomas-Fermi expression for the
kinetic part. OFMD simulations are numerically available for
temperatures up to 1000 eV, where QMD simulations became
prohibitive. The electronic density, which is the only variable in
OFMD simulations, is Fourier transformed during the calculation, and
numerical convergence in terms of the mesh size has been insured for
all the simulations.

Two nondimensional parameters are used to characterize the state of plasmas,
namely, the ion-ion coupling and the electron degeneracy parameters. The
former one is commonly defined as $\Gamma_{ii}=Z^{\ast2}/(k_{B}Ta)$, which
describes the ratio of the mean electrostatic potential energy and the mean
kinetic energy of the ions. Here, $Z^{\ast}$ is the average ionization degree,
which is equal to 2 in the present system and $a$ is the ionic sphere radius.
The degeneracy parameter $\theta=T/T_{F}$ is the ratio of the temperature to
the Fermi temperature $T_{F}=(3/\pi^{2}n_{e})^{2/3}/3$. Both of the parameters
are summarized in Table \ref{twoparameter} for the studied densities and temperatures.

\begin{table}[tbh]
\caption{Ion-ion coupling parameter ($\Gamma_{ii}$) and electron degeneracy
parameter $\theta$.}%
\centering
\begin{tabular}
[c]{ccccccccccccccccccccccccccccccccccccc}\hline\hline
\multicolumn{1}{r}{$\rho$ (g/cm$^{3}$)} & \multicolumn{2}{c}{400} &
\multicolumn{2}{c}{480} & \multicolumn{2}{c}{600} & \multicolumn{2}{c}{800} &
&  &  &  &  &  &  &  &  &  &  &  &  &  &  &  &  &  &  &  &  &  &  &  &  &  &
& \\\hline
\multicolumn{1}{l}{T (eV)} & $\Gamma_{ii}$ & $\theta$ & $\Gamma_{ii}$ &
$\theta$ & $\Gamma_{ii}$ & $\theta$ & $\Gamma_{ii}$ & $\theta$ &  &  &  &  &
&  &  &  &  &  &  &  &  &  &  &  &  &  &  &  &  &  &  &  &  &  &  & \\\hline
10 & 28.85 & 0.02 & 30.66 & 0.02 & 33.03 & 0.01 & 36.35 & 0.01 &  &  &  &  &
&  &  &  &  &  &  &  &  &  &  &  &  &  &  &  &  &  &  &  &  &  &  & \\
20 & 14.43 & 0.04 & 15.33 & 0.03 & 16.51 & 0.03 & 18.18 & 0.02 &  &  &  &  &
&  &  &  &  &  &  &  &  &  &  &  &  &  &  &  &  &  &  &  &  &  &  & \\
50 & 5.77 & 0.09 & 6.13 & 0.08 & 6.61 & 0.07 & 7.27 & 0.06 &  &  &  &  &  &  &
&  &  &  &  &  &  &  &  &  &  &  &  &  &  &  &  &  &  &  &  & \\
100 & 2.89 & 0.18 & 3.07 & 0.16 & 3.30 & 0.14 & 3.64 & 0.11 &  &  &  &  &  &
&  &  &  &  &  &  &  &  &  &  &  &  &  &  &  &  &  &  &  &  &  & \\
200 & 1.44 & 0.36 & 1.53 & 0.32 & 1.65 & 0.27 & 1.82 & 0.22 &  &  &  &  &  &
&  &  &  &  &  &  &  &  &  &  &  &  &  &  &  &  &  &  &  &  &  & \\
300 & 0.96 & 0.54 & 1.02 & 0.47 & 1.10 & 0.41 & 1.21 & 0.34 &  &  &  &  &  &
&  &  &  &  &  &  &  &  &  &  &  &  &  &  &  &  &  &  &  &  &  & \\
400 & 0.72 & 0.71 & 0.77 & 0.63 & 0.83 & 0.54 & 0.91 & 0.45 &  &  &  &  &  &
&  &  &  &  &  &  &  &  &  &  &  &  &  &  &  &  &  &  &  &  &  & \\
500 & 0.58 & 0.89 & 0.61 & 0.79 & 0.66 & 0.68 & 0.73 & 0.56 &  &  &  &  &  &
&  &  &  &  &  &  &  &  &  &  &  &  &  &  &  &  &  &  &  &  &  & \\
800 & 0.36 & 1.43 & 0.38 & 1.26 & 0.41 & 1.09 & 0.45 & 0.90 &  &  &  &  &  &
&  &  &  &  &  &  &  &  &  &  &  &  &  &  &  &  &  &  &  &  &  &
\\\hline\hline
\end{tabular}
\label{twoparameter}%
\end{table}

\begin{figure}[pt]
\centering
\includegraphics[width=8.0cm]{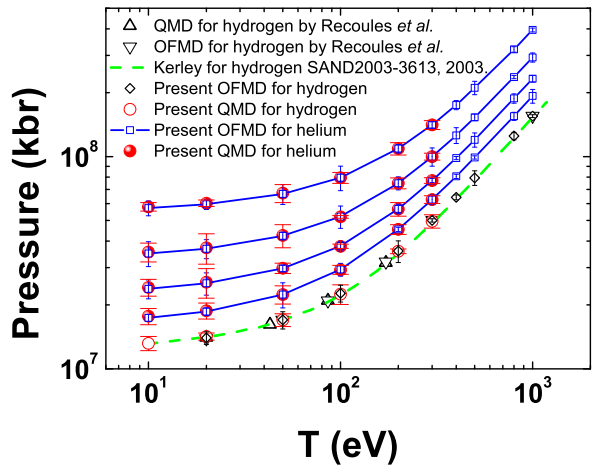}\caption{(Color online) Calculated EOS
(pressure versus temperature) for dense helium. The QMD and OFMD
results, where the error bars are shown by using wide and narrow
caps, are labelled as red filled circles and blue open squares,
respectively. Previous theoretical predictions obtained by Recoules
\emph{et al.} (up and down triangles) \cite{PBX:Recoules:2009} and
Kerley (green dashed line) \cite{PBX:Kerley:2003}, where the EOS of
dense hydrogen along the 80 g/cm$^{3}$ isochore, are also
shown for comparison.}%
\label{fig_eos}%
\end{figure}

The accurate determination of the transport coefficients depends on
a precise description of the EOS. The wide-range EOS are obtained
from both quantum and semiclassical molecular dynamic simulations.
As have been shown in Table \ref{twoparameter}, our simulations
start from strongly coupled and highly degenerate states, then reach
moderate coupling and partial degeneracy states. To examine our
results, we also calculate the EOS of hydrogen along 80 g/cm$^{3}$
isochore up to 1000 eV, and the data are in accordance with previous
theoretical predictions \cite{PBX:Recoules:2009,PBX:Kerley:2003}, as
shown in Fig. \ref{fig_eos}. Good agreements are also shown between
our QMD and OFMD results, due to the ideal metallization of helium
in the hot dense regime. The simulated EOS along temperature for
dense helium shows systematic behavior, and smooth functions
($P=\sum A_{ij}\rho^{i}T^{j}$), which can be simply used in
hydrodynamic simulations for hot dense helium and astrophysical
applications, have been constructed (coefficients $A_{ij}$ have been
listed in Table \ref{coefficient_P}).

\begin{table}[tbh]
\caption{Pressure (kbar) expansion coefficients $A_{ij}$ in terms of density
(g/cm$^{3}$) and temperature (eV).}%
\centering
\begin{tabular}
[c]{cccccc}\hline\hline
$i$ & $A_{i0}$ & $A_{i1}$ & $A_{i2}$ &  & \\\hline
0 & 26.77 & 2710.10 & -7.02 &  & \\
1 & 10819.04 & 398.98 & 0.08 &  & \\
2 & 69.92 & -0.07 & 0.00 &  & \\\hline\hline
\end{tabular}
\label{coefficient_P}%
\end{table}

The linear response of hot dense helium to external electrical field
and temperature gradient can be characterized by the electrical and
heat current densities. The key to evaluate these linear-response
transport properties is the kinetic coefficients based on the
Kubo-Greenwood formula
\begin{equation}
\hat{\sigma}(\epsilon)=\frac{1}{\Omega}\sum_{k,k^{\prime}}|\langle\psi
_{k}|\hat{v}|\psi_{k^{\prime}}\rangle|^{2}\delta(\epsilon_{k}-\epsilon
_{k^{\prime}}-\epsilon),\label{kubo}%
\end{equation}
where $\langle\psi_{k}|\hat{v}|\psi_{k^{\prime}}\rangle$ are the velocity
matrix elements, $\Omega$ is the volume of the supercell, and $\epsilon_{k}$
are the electronic eigenvalues. The kinetic coefficients $\mathcal{L}_{ij}$ in
the Chester-Thellung version \cite{PBX:Chester:1961} are given by
\begin{equation}
\mathcal{L}_{ij}=(-1)^{i+j}\int d\epsilon\hat{\sigma}(\epsilon)(\epsilon
-\mu)^{(i+j-2)}(-\frac{\partial f(\epsilon)}{\partial\epsilon}%
),\label{coefficient}%
\end{equation}
with $\mu$ being the chemical potential and $f(\epsilon)$ the Fermi-Dirac
distribution function. We obtain the electrical conductivity $\sigma$
\begin{equation}
\sigma=\mathcal{L}_{11},\label{sigma}%
\end{equation}
and electronic thermal conductivity $K$ is
\begin{equation}
K=\frac{1}{T}(\mathcal{L}_{22}-\frac{\mathcal{L}_{12}^{2}}{\mathcal{L}_{11}%
}),\label{thermal}%
\end{equation}
where $T$ is the temperature. Eqs. (\ref{sigma}) and (\ref{thermal})
are energy-dependent, then the electrical conductivity and
electronic thermal conductivity are obtained through extrapolating
to zero energy. Those formulation are implemented in the ABINIT
code, and have lead to good results for liquid aluminum
\cite{PBX:Recoules:2005} and hot dense hydrogen
\cite{PBX:Recoules:2009}, where Troullier-Martins potential and
Coulombic potential were used respectively. Within the framework of
finite-temperature DFT, chemical potential is evaluated through
fitting the set of occupation numbers corresponding to the set of
eigenvalues with the usual functional form for the Fermi-Dirac
distribution. The $\delta$ function in Eq. \ref{kubo} is broaden in
the calculations by a Gaussian function, which has been tested to
obtain smooth curves.

The Lorentz number $L$ is defined as
\begin{equation}
L=\frac{K}{\sigma T}=\gamma\frac{e^{2}}{k_{B}^{2}}.\label{lorentz}%
\end{equation}
The rule of force responsible for the electronic scattering characterizes
$\gamma$. From the Wiedemann-Franz law, $\gamma$ is $\pi^{2}/3$ and $L$ is
$2.44\times10^{-8}$ in the degenerate regime (low temperature). For the
nondegenerate case (high temperature), $\gamma$ is 1.5966 and $L$ is
$1.18\times10^{-8}$. No classical assumptions are available for $\gamma$ value
in the intermediate region. As a consequence, the electronic thermal
conductivity can not be deduced from the electrical conductivity by using
Wiedemann-Franz law. Direct predictions of electrical and thermal conductivity
from QMD simulations are then rather interesting.

In order to get converged transport coefficients, ten independent snapshots,
which are selected during one molecular dynamics simulation at given
conditions, are picked up to calculate electrical conductivity and electronic
thermal conductivity as running averages. For temperatures below 300 eV,
atomic configurations are directly extracted from QMD simulations, while they
are taken from OFMD simulations for higher temperatures. Let us stress here
that, one-body electronic states are populated according to a Fermi-Dirac
distribution in QMD simulations, and a large number of occupied orbitals are
introduced at high temperatures. Thus this method is rather time consuming.

\begin{figure}[pt]
\centering
\includegraphics[width=8.0cm]{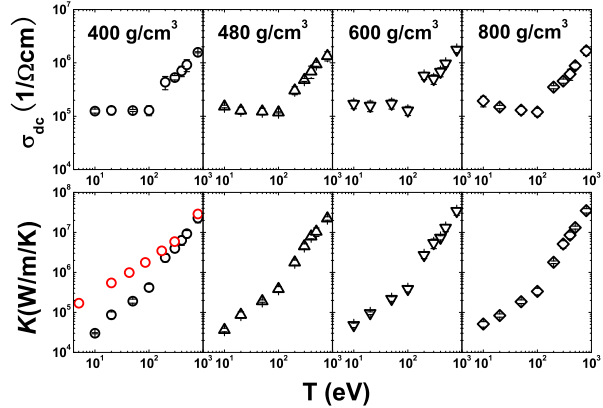}\caption{(Color online) Calculated electrical
conductivity (upper panels) and electronic thermal conductivity
(lower panels) as functions of temperature at four densities of 400,
480, 600, and 800 g/cm$^{3}$. For comparison, electronic thermal
conductivity for dense
hydrogen is also shown as red open circles.}%
\label{fig_cond}%
\end{figure}

In Fig. \ref{fig_cond} we plot the electrical conductivity and
electronic thermal conductivity as functions of temperature. While
at relatively low temperatures up to $100$ eV the electrical
conductivity changes slowly, it turns to smoothly increase at higher
temperatures. The electronic thermal conductivity shows a sustaining
increase with temperature, as in Fig. \ref{fig_cond} (lower panel).
As temperatures arise from 10 to 800 eV, electronic thermal
conductivity is of great importance in ICF applications, where these
thermodynamic conditions are encountered in the pellet target
fusion. In general ICF designs, several models are used to evaluate
thermal conductivity, such as Hubbard and Lee-More models
\cite{PBX:Hubbard:1966,PBX:Lee:1984}. Previous theoretical
predictions for dense hydrogen have demonstrated that the electronic
thermal conductivity obtained by QMD simulations are in fairly good
agreement with the Hubbard model. For very high temperatures, the
results for Lee-More model merge into the Spitzer thermal
conductivity \cite{PBX:Spitzer:1953}, which are accordant with QMD
simulations \cite{PBX:Recoules:2009}. As a comparison, we also plot
the thermal conductivity for dense hydrogen in Fig. \ref{fig_cond}.
It is indicated that at temperatures below 200 eV, electronic
thermal conductivity for hydrogen plasma is larger than that of
helium. However, they tend to merge with each other for higher
temperatures.

\begin{figure}[pt]
\centering
\includegraphics[width=8.0cm]{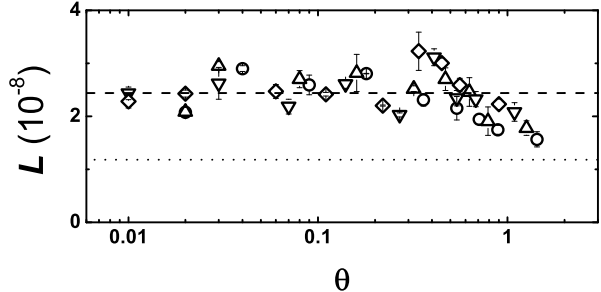}\caption{Lorenz number as a function of
electron degeneracy parameter. To distinguish different densities,
open circle, up triangle, down triangle, and open diamond are used
to denote densities of 400, 480, 600, 800 g/cm$^{3}$, respectively.
The relative degeneracy parameter can be read from Table.
\ref{twoparameter}. The dashed line is the value
of the degenerate limit, and the dotted line is the value of nondegenerate case.}%
\label{fig_Lorenz}%
\end{figure}

Using the calculated electrical and thermal conductivity, the
Lorentz ratio is extracted and shown in Fig. \ref{fig_Lorenz}. In
the strong coupling and degenerate region, the computed Lorentz
number vibrates around ideal value of 2.44$\times10^{-8}$, which is
predicted by the nearly free electron model. This model is valid in
the case where Born approximation is applicable because scattering
of the electrons by the ions is sufficiently weak. With the increase
of temperature, where dense helium enters moderate coupling and a
partial degeneracy regime, Lorenz number tends to decrease and
approach to the nondegenerate value of 1.18$\times10^{-8}$.

In summary, we have performed \emph{ab initio} QMD simulations to
study the thermophysical properties of dense helium plasmas under
extreme conditions as reached in ICF experiments. As a result,
highly converged EOS data (pressures up to $10^{8}$ kbar), for which
Coulombic potential with a very small cutoff radius is adopted to
model ultra-dense helium, have been obtained. We have constructed
smooth functions to fit the QMD data for the pressure, which is
applicable for astrophysics and ICF designs. Using Kubo-Greenwood
formula, the electrical conductivity and electronic thermal
conductivity have been systematically determined and carefully
discussed. The examination of Lorenz number, as well as that of the
calculated plasmas parameters (Table I), has indicated a gradual
transition from a strong coupling and degenerate state to a moderate
coupling and partial degeneracy regime for dense helium. We expect
the present simulated results provide a guiding line in the
practical ICF hydrodynamical simulations with the helium plasma
participated.

This work was supported by NSFC under Grants No. 11005012 and No. 51071032, by
the National Basic Security Research Program of China, and by the National
High-Tech ICF Committee of China.


\begin{thebibliography}{99}                                                                                               %
\bibitem {PBX:Atzeni:2004}S. Atzeni and J. Meyer-ter-Vehn, \emph{The Physics
of Inertial Fusion: Beam Plasma Interaction, Hydrodynamics, Hot Dense Matter},
International Series of Monographs on Physics (Clarendon Press, Oxford, 2004).

\bibitem {PBX:Lindl:1998}J. D. Lindl, \emph{Inertial Confinement Fusion: The
Quest for Ignition and Energy Gain Using Indirect Drive}, (Springer- Verlag,
New York, 1998).

\bibitem {PBX:Nuckolls:1972}J. Nuckolls \emph{et al.}, Nature (London)
\textbf{239} 139 (1972).

\bibitem {PBX:Seka:2008}W. Seka \emph{et al.}, Phys. Plasmas \textbf{15}
056312 (2008).

\bibitem {PBX:Boehly:2006}T. R. Boehly \emph{et al.}, Phys. Plasmas
\textbf{13} 056303 (2006).

\bibitem {PBX:Azechi:1997}H. Azechi \emph{et al.}, Phys. Plasmas \textbf{4}
4079 (1997).

\bibitem {PBX:Glendinning:1997}S. G. Glendinning \emph{et al.}, Phys. Rev.
Lett. \textbf{78} 3318 (1997).

\bibitem {PBX:Lobatchev:2000}V. Lobatchev \emph{et al.}, Phys. Rev. Lett.
\textbf{85} 4522 (2000).

\bibitem {PBX:Lindl:2004}J. D. Lindl \emph{et al.}, Phys. Plasmas \textbf{11}
339 (2004).

\bibitem {PBX:Cai:2009}H. Cai \emph{et al.}, Phys. Rev. Lett. \textbf{102}
245001 (2009).

\bibitem {PBX:Spitzer:1953}L. Spitzer \emph{et al.}, Phys. Rev. \textbf{89}
977 (1953).

\bibitem {PBX:Hubbard:1966}W. B. Hubbard \emph{et al.}, Astrophys. J.
\textbf{146} 858 (1966).

\bibitem {PBX:Brysk:1975}H. Brysk \emph{et al.}, Plasma Phys. \textbf{17} 473 (1975).

\bibitem {PBX:Lee:1984}Y. T. Lee \emph{et al.}, Phys. Fluids \textbf{27} 1273 (1984).

\bibitem {PBX:Kietzmann:2008}A. Kietzmann \emph{et al.}, Phys. Rev. Lett.
\textbf{101} 070401 (2008).

\bibitem {PBX:Lorenzen:2009}W. Lorenzen \emph{et al.}, Phys. Rev. Lett.
\textbf{102} 115701 (2009).

\bibitem {PBX:abinit}The ABINIT code, which is a common project of the
Universit\'{e} Catholique de Louvain, Corning Incorporated, Commissariat \`{a}
l'Energie Atomique, Universit\'{e} de Li\`{e}ge, Mitsubishi Chemical Corp. and
other contributions, is available at http://www.abinit.org.

\bibitem {PBX:Gonze:2002}X. Gonze \emph{et al.}, Comput. Mater. Sci.
\textbf{25} 478 (2002).

\bibitem {PBX:Bottin:2008}F. Bottin \emph{et al.}, Comput. Mater. Sci.
\textbf{42} 329 (2008).

\bibitem {PBX:Goedecker:1996}S. Goedecker \emph{et al.}, Phys. Rev. B
\textbf{54} 1703 (1996).

\bibitem {PBX:timestep}The time steps have been taken as $\Delta
t=a/20\sqrt{k_{B}T/m_{He}}$, where $a=(3/4\pi n_{i})^{1/3}$ is the ionic
sphere radius ($n_{i}$ is the ionic number density), $k_{B}T$ presents the
kinetic energy, and $m_{He}$ is the ionic mass. A few hundred of plasma
periods are included in this choice of simulations.

\bibitem {PBX:Brack:2003}M. Brack and R. K. Bhaduri, \emph{Semiclassical
Physics} (Westview Press, Boulder, CO, 2003), ISBN 0813340845.

\bibitem {PBX:Recoules:2009}V. Recoules, \emph{et al.}, Phys. Rev. Lett.
\textbf{102} 075002 (2009).

\bibitem {PBX:Kerley:2003}G. I. Kerley, Sandia National Laboratory Tech. Rep.
SAND2003-3613, 2003.

\bibitem {PBX:Chester:1961}G.V. Chester and A. Thellung, Proc. Phys. Soc.
(London) 77, 1005 (1961).

\bibitem {PBX:Recoules:2005}V. Recoules, \emph{et al.}, Phys. Rev. B
\textbf{72} 104202 (2005).
\end{thebibliography}

\end{document}